\documentclass[12pt]{article}

\usepackage{amsfonts}
\begin{document}

\def\ba{\begin{eqnarray}}
\def\ea{\end{eqnarray}}

\begin{titlepage}
\title{DIRAC EQUATION IN SPACETIMES \\WITH NON-METRICITY AND TORSION}
\author{ M. Adak  \\
 {\small Department of Physics, Pamukkale University,}\\
{\small 20100 Denizli, Turkey} \\ {\small madak@pamukkale.edu.tr} \\ \\
T. Dereli \\
 {\small Department of Physics, Ko\c{c} University,}\\
 {\small 80910 Sar{\i}yer, \.{I}stanbul, Turkey  }  \\
{\small tdereli@ku.edu.tr} \\   \\
L. H. Ryder\\
{\small Department of Physics, University of Kent,}\\
{\small  Canterbury, Kent CT2 7NF, UK}\\
{\small l.h.ryder@ukc.ac.uk}}
\vskip 1cm
\maketitle

\begin{abstract}

\noindent  Dirac equation is written in a non-Riemannian
spacetime with torsion and non-metricity by lifting the
connection from the tangent bundle to the spinor bundle over
spacetime. Foldy-Wouthuysen transformation of the Dirac equation
in a Schwarzschild background spacetime is considered and it is
shown that both the torsion and non-metricity couples to the
momentum and spin of a massive, spinning  particle. However, the
effects are small to be observationally significant.

\end{abstract}
\end{titlepage}

\section{Introduction}

Einstein's general relativity provides an elegant
(pseudo-)Riemannian formulation of gravitation in the absence of
matter. In the variational approach, Einstein's field equations
are obtained by considering variations of the Einstein-Hilbert
action with respect to the metric and its associated Levi-Civita
connection of the spacetime. That is, the absence of matter means
that the connection is metric compatible and torsion free, a
situation which is natural but not always convenient. A number of
developments in physics in recent years suggest the possibility
that the treatment of spacetime might involve more than a
Riemannian structure:
\begin{itemize}
  \item  The vain effort to quantize gravity with standard field
         theoretic methods. This is perhaps so far the strongest
         piece of evidence for going beyond a
         geometry dominated by classical metric concepts.
  \item  The generalization of the theory of elastic continua to
         4-dimensional space-time. This may provide a
         physical interpretation of the non-Riemannian structures
         which emerge in the theory.
  \item  The study of the early universe, in particular
         singularity theorems, gave us clues for solving the problem
         of unification of all
         interactions and the related problem of compactification
         of dimensions. Inflationary models with
         dilaton-induced Weyl co-vector is another problem where
         we meet non-Riemannian structures.
  \item  The description of hadronic (or nuclear) matter in terms
         of extended structures. The string theories are the best
         examples.
\end{itemize}

In the framework of string theories there are hints that by using
non-Riemannian geometry we may accommodate several degrees of
freedom coming from the low energy limit of string interactions in
terms of a connection with non-metricity and
torsion~\cite{der4}. It is interesting to observe that, since
string theories are expected to produce effects which are at least
in principle testable at low energies, there may be chances to
obtain non-Riemannian models with predictions that can be tested.

Although theories in which the non-Riemannian geometrical fields
are dynamical in the absence of matter are more elusive to
interpret, they may play an important role in certain
astrophysical contexts. Part of the difficulty in interpreting
such fields is that there is little experimental guidance
available for the construction of a viable theory that can compete
effectively with general relativity in domains that are currently
accessible to observation. In such circumstances one must be
guided by the classical solutions admitted by theoretical models
that admit dynamical non-Riemannian
structures~\cite{ben1,heh,der2}.

In the next section we give some basic concepts in
differential geometry and mathematical tools in order to perform
the calculations given later in the paper. In
section~\ref{sec.dirac} we write down  the Dirac equation
in non-Riemannian spacetimes with non-vanishing curvature, torsion
and non-metricity by the Kosmann lift of the
connection given in the frame bundle to the spinor
bundle ~\cite{kos,hur,fat}.
The physical meaning of curvature is well-known. In
order to gain insight concerning torsion and  non-metricity we
investigate the low energy limit of the Dirac equation~\cite{ryd}.
While doing this we perform the Foldy-Wouthuysen
transformation~\cite{fol,nik,obu}; thus uncoupling the positive and the
negative energy states of the Dirac equation.
The final section is reserved for discussion.

\section{Non-Riemannian spacetime}

Spacetime is denoted by the triple $ \{M,g,\nabla \} $ where M is
a 4-dimensional differentiable manifold, equipped with a
Lorentzian metric $ g $ which is a second rank, covariant,
symmetric, non-degenerate tensor and $ \nabla $ is a linear connection
which defines parallel transport of vectors (or more generally
tensors). With an orthonormal basis $\{ X_a \}$, 
\ba 
g = \eta_{ab}e^a \otimes e^b \;\;\; , \;\; a,b,\cdots = 0,1,2,3
\label{metric}
\ea
where $\eta_{ab}$ is the Minkowski metric which is a matrix whose
diagonal terms are $-1,1,1,1$ and $\{ e^a \}$ is the orthonormal
co-frame. The local orthonormal frame $\{ X_a \}$ is dual to the
co-frame $\{e^a \}$;
 \ba
   e^b(X_a)=\delta^b_a \; . \ea
The spacetime orientation is set by the choice
$\epsilon_{0123}=+1 $. In addition, the connection is specified by
a set of connection 1-forms ${\Lambda^a}_b$. The non-metricity  
1-forms, torsion 2-forms and curvature 2-forms are defined
through  the Cartan structure equations 
\ba
2Q_{ab} &:=& -D\eta_{ab} = \Lambda_{ab} +\Lambda_{ba} \, ,\label{nonmet}\\
T^a &:=& De^a = de^a + {\Lambda^a}_b \wedge e^b \, , \label{torsion}\\
{R^a}_b &:=& D{\Lambda^a}_b := d{\Lambda^a}_b +{\Lambda^a}_c
\wedge {\Lambda^c}_b \; . \label{curva}
\ea 
d, D, $\imath_a$, * denote the exterior derivative,
the covariant exterior derivative, the interior derivative and the
Hodge star operator, respectively.
The linear
connection 1-forms can be decomposed uniquely as
follows~\cite{der2}
 \ba
  {\Lambda^a}_b = {\omega^a}_b + {K^a}_b
          + {q^a}_b + {Q^a}_b \;  \label{connec}
 \ea
where $ {\omega^a}_b $ are the  Levi-Civita connection 1-forms that satisfy
 \ba
     de^a + {\omega^a}_b \wedge e^b = 0 \; , \label{levi}
 \ea
$ {K^a}_b $ are the contortion 1-forms such that
 \ba
     {K^a}_b \wedge e^b = T^a \; , \label{contor}
 \ea
and $ {q^a}_b $ are the anti-symmetric tensor 1-forms defined by
 \ba
     q_{ab} = -(\imath_a Q_{bc}) \wedge e^c
        + (\imath_b Q_{ac}) \wedge e^c \, . \label{antisy}
 \ea
 In the above  decomposition the symmetric part
 \ba
    \Lambda_{(ab)} = Q_{ab} \label{symm}
 \ea
while the anti-symmetric part
 \ba
  \Lambda_{[ab]} = \omega_{ab} + K_{ab} + q_{ab} \; . \label{asymm}
 \ea

It is cumbersome to take into account all components of
non-metricity and  torsion  in gravitational models.
Therefore we will be content with dealing  only with certain
irreducible parts of them
 to gain physical insight. The irreducible
decompositions of torsion and non-metricity  invariant
under the Lorentz group are summarily given below. For details one may consult Ref.~\cite{heh}.
The non-metricity 1-forms $ Q_{ab} $ can be split into their
trace-free $ \overline{Q}_{ab} $ and the trace parts as
 \ba
    Q_{ab} = \overline{Q}_{ab} + \eta_{ab}Q          \label{118}
 \ea
where the Weyl 1-form $Q={Q^a}_a$ and $ \eta^{ab}\overline{Q}_{ab} = 0 $. Let us define
 \ba
    \Lambda_b &:=& \imath_a { \overline{Q}^a}_b, \;\;\;\;\;\;\;\;\;\;
                   \Lambda := \Lambda_a e^a,    \nonumber  \\
    \Theta_b &:=& {}^*(\overline{Q}_{ab} \wedge e^a), \;\;\;
    \Theta := e^b \wedge \Theta_b, \;\;\;
    \Omega_a := \Theta_a -\frac{1}{3}\imath_a\Theta    \label{119}
 \ea
as to use them in the decomposition of $ Q_{ab} $ as
 \ba
     Q_{ab} = Q_{ab}^{(1)} + Q_{ab}^{(2)} +
            Q_{ab}^{(3)} +Q_{ab}^{(4)}             \label{120}
 \ea
where
 \ba
    Q_{ab}^{(2)} &=& \frac{1}{3} {}^*(e_a \wedge \Omega_b +e_b
    \wedge \Omega_a)   \\
    Q_{ab}^{(3)} &=& \frac{2}{9}( \Lambda_a e_b +\Lambda_b e_a
    -\frac{1}{2} \eta_{ab} \Lambda )  \\
    Q_{ab}^{(4)} &=& \frac{1}{4} \eta_{ab} Q    \\
    Q_{ab}^{(1)} &=& Q_{ab}-Q_{ab}^{(2)}
                 - Q_{ab}^{(3)} -Q_{ab}^{(4)} \;.
 \ea
We have $ \imath^a Q_{ab}^{(1)} =\imath^a Q_{ab}^{(2)} =0,
\;\;\;\;\;  \eta^{ab}Q_{ab}^{(1)} = \eta^{ab}Q_{ab}^{(2)}
=\eta^{ab}Q_{ab}^{(3)} = 0, \;\;$ and
$ e^a \wedge Q_{ab}^{(1)} =0 $.
In a similar way the  irreducible decomposition of $ T^a $'s invariant under the Lorentz group are given in terms of
 \ba
   \alpha =\imath_a T^a \; , \;\;\; \sigma = e_a \wedge T^a
 \ea
so that
 \ba
    T^a = {T^a}^{(1)} +{T^a}^{(2)} +{T^a}^{(3)}      \label{121}
 \ea
where
 \ba
    {T^a}^{(2)} &=& \frac{1}{3}e^a \wedge \alpha \; , \\
    {T^a}^{(3)} &=& \frac{1}{3} \imath^a \sigma \; , \\
    t^a &:=&{T^a}^{(1)} = T^a-{T^a}^{(2)} -{T^a}^{(3)} \; .
 \ea
Here $ \imath_a t^a = \imath_a {T^a}^{(3)} = 0, \;\;\; e_a
\wedge t^a = e_a \wedge {T^a}^{(2)} = 0 $.
To give the
contortion components in terms of the irreducible components of
torsion, we firstly  write
 \ba
     2K_{ab} = \imath_a T_b - \imath_b T_a
     - (\imath_a \imath_b T_c)e^c             \label{122}
 \ea
from (\ref{contor}) and then substituting (\ref{121}) into above
we find
 \ba
   2K_{ab} &=& \imath_a t_b - \imath_b t_a
               - (\imath_a \imath_b t_c)e^c \nonumber \\
           &{}& +\frac{2}{3} ( e_a \wedge \imath_b \alpha
                - e_b \wedge \imath_a \alpha )   \nonumber \\
           &{}& + \frac{2}{3} ( \imath_a \imath_b \sigma )
 - \frac{1}{3} ( \imath_a \imath_b \imath_c \sigma )e^c.  \label{123}
 \ea
In components $ K_{ab} = K_{c,ab}e^c \; , \;\;\;  t_a =
\frac{1}{2}t_{bc,a}e^{bc} \; , \;\;\; \alpha = F_a e^a \; ,
\;\;\; \sigma = \frac{1}{3!} \sigma_{abc} e^{abc} $ this becomes
 \ba
    K_{c,ab} &=& \frac{1}{2} ( t_{ab,c}
                 - t_{bc,a} + t_{ac,b} ) \nonumber \\
           &{}& +\frac{1}{3}( F_b \eta_{ac}
         - F_a \eta_{bc})  -\frac{1}{6} \sigma_{abc}. \label{D.15}
 \ea

\section{Dirac equation}\label{sec.dirac}

Since gauge theories seem important for the description of
fundamental interactions it appears natural to exploit any gauge
structure present in theories of gravity. Different authors,
however, adopt different criteria in order to determine what
properties a theory should possess in order for it to qualify as a
gauge theory. We take the gravitational gauge group to be the
local Lorentz group~\cite{ben1} and do calculations in terms
of $ { \mathbf{Spin}}_+( 3,1 ) $ structure group, that is the
double covering of Lorentz group, keeping in mind the inclusion of
spin structure. We will work with exterior forms taking their
values in the $ { {\mathcal{C}}}\ell_{3,1} $ Clifford algebra. $ {
{\mathcal{C}}}\ell_{3,1} $ with an orthonormal basis $ \{
\gamma_0, \gamma_1, \gamma_2, \gamma_3 \} $ is generated by the
relations
 \ba
    \gamma^a \gamma^b + \gamma^b \gamma^a = 2 \eta^{ab}I_{4\times 4}
 \ea
where we use the following Dirac matrices
 \ba
   \gamma_0  &=&  i \left (
                 \begin{array}{cc}
                                   -I & 0  \\
                                   0  & I
                 \end{array}
          \right )             \; , \;\;\;
   \gamma_1  =  i \left (
                 \begin{array}{cc}
                                   0    &  \sigma^1  \\
                                  -\sigma^1 &  0
                 \end{array}
          \right ) \nonumber \\
  \gamma_2  &=&  i \left (
                 \begin{array}{cc}
                                   0     &  \sigma^2  \\
                                   -\sigma^2 &  0
                 \end{array}
          \right )  \; , \;\;\;
  \gamma_3  =  i \left (
                 \begin{array}{cc}
                                  0     & \sigma^3  \\
                                   -\sigma^3 & 0
                 \end{array}
          \right )  \; . \label{dirmat}
 \ea
Here $\sigma^i $ are the Pauli matrices.

A 4-component Dirac spinor is a section of the spinor bundle over
spacetime  whose
fibers are  ${\mathbb{C}}^4$. This bundle is simply the vector bundle
associated to the tangent bundle. Then a  connection
in the tangent bundle is lifted to the spinor  bundle via the
Kosmann lift~\cite{kos,hur,fat}
 \ba
   \Omega = \frac{1}{4} \Lambda^{ab}\gamma_a \gamma_b \label{Omega}
 \ea
where
 \ba
    \gamma_a \gamma_b = \eta_{ab} + 2 \sigma_{ab}
 \ea
is the Clifford product in the spinor bundle. The anti-symmetric part of the Clifford product
 \ba
     \sigma_{ab} = \frac{1}{4} [ \gamma_a, \gamma_b]
 \ea
carries a representation of the Lorentz group  and
${\Lambda^a}_b$ is the connection 1-form in the tangent space.
The covariant exterior derivative of the Dirac
spinor may therefore be written explicitly as
 \ba
  D\psi = d\psi + \frac{1}{2}\Lambda^{[ab]}\sigma_{ab} \psi
             + \frac{1}{4}Q \psi \; .\label{covder}
 \ea
It should be noted  that the spin connection $ \Omega $ depends on the
non-metricity tensor via $ q_{ab} $ and the Weyl 1-form $Q$.
Thus the {\bf Dirac equation } in a non-Riemannian spacetime is
given by
 \ba
   ^{*}\gamma \wedge D \psi + \mu^{*}1 \psi = 0 \label{direqn}
 \ea
where the Clifford algebra ${\mathcal{C}}\ell_{3,1}$-valued 1-form
$\gamma = \gamma^a e_a $ and  $\mu =\frac{mc}{\hbar}$.

\subsection{ A  spinning particle in a non-Riemannian
               spacetime }

In this section we investigate the motion of a massive, spinning
particle in a
non-Riemannian spacetime with curvature, torsion and certain
components of non-metricity. In order to do that we construct the
Dirac Hamiltonian of the particle. This will be done here in the
background Schwarzschild metric of a heavy,
static body of mass $M$ such as the Sun. In isotropic
coordinates $(t,x,y,z)$ the  metric is
given by\footnote{We use a unit system such that $c=G=1$.}
 \ba
    g= -f^2 dt^2 + g^2 (dx^2 +dy^2 +dz^2 )          \label{h1}
 \ea
where the metric functions $f$ and $g$ are functions of $ r^2
=x^2 +y^2 +z^2$ only.
The Schwarzschild geometry is specified by the functions
 \ba
   f &=& \frac{1 + \xi}{1 - \xi} \; ,  \\
   g &=&  (1-\xi)^2 \; ,
 \ea
where $\xi =-\frac{M }{2 r }\;$.
 The orthonormal co-frame may be cast as
 \ba
  e^0 = f dt \; , \;\;\;\;\;\; e^i = g dx^i \; \;, \;\;\; i= 1,2,3 \;. \label{h2}
 \ea
Then the Levi-Civita connection 1-forms are calculated from
(\ref{levi}) as
 \ba
     \omega_{0i} = -\frac{(\partial_i f)}{fg}e^0 \; ,\;\;\;
 \omega_{ij} = \frac{1}{g^2} [ (\partial_j g)e_i
                      - (\partial_i g)e_j ] .\label{h3}
 \ea
Therefore the  covariant exterior derivative of the Dirac spinor
$\psi$  becomes
 \ba
    D\psi &=& e^0 [ \frac{(\partial_t \psi)}{f}
    - \frac{1}{4}Q^0 \psi
    - \frac{(\partial_i f)}{fg}\sigma^{0i} \psi ] \nonumber \\
     & &  + e^i [ \frac{(\partial_i \psi)}{g} + \frac{1}{4} Q_i
    + \frac{(\partial_j g)}{g^2}\sigma^{ij} \psi ] \nonumber \\
    & & + e^c \frac{1}{2} \Omega_{c,ab}\sigma^{ab}\psi           \label{h4}
 \ea
where $ \Omega_{ab} =K_{ab} +q_{ab} $ and
the Weyl 1-form $Q = Q_a e^a $. 
After substituting (\ref{h4}) into (\ref{direqn}) and 
using $ {}^* e^f \wedge e^a =-\eta^{fa}{}^*1 $ we obtain
 \ba
    \frac{\gamma^0}{f} \frac{\partial}{\partial t}&=& \mu \psi
    + \gamma^0 \frac{1}{4}Q^0  \psi \nonumber \\
    & &  - \gamma^i[\frac{(\partial_i \psi)}{g}
    + \frac{1}{4}Q^i  + \frac{(\partial_j g)}{g^2}
       \sigma^{ij} \psi ]{} \nonumber \\
    & & + \frac{(\partial_i f)}{fg}\gamma^0 \sigma^{0i}\psi
    - \frac{1}{2} \Omega_{c,ab}\gamma^c\sigma^{ab} \psi .      \label{h5}
 \ea
Now using the identity, with $\gamma_5 = \gamma^0 \gamma^1
\gamma^2 \gamma^3 $,
 \ba
    \gamma^a \sigma^{bc} = \frac{1}{2}\eta^{ab}\gamma^c
    -\frac{1}{2}\eta^{ac}\gamma^b -\frac{1}{2}
    \epsilon^{abcd}\gamma_5 \gamma_d   \label{iden}
 \ea
we rearrange  this as
 \ba
    \frac{1}{f}\frac{\partial \psi}{\partial t} &=& \frac{\gamma^0 \gamma^i}{g}
 [\partial_i + \frac{g}{4}Q^i+ \frac{\partial_i(fg)}{2fg} ]\psi
   \nonumber \\
 & &  +\frac{1}{2} {\Omega^{a,}}_{ab} \gamma^0 \gamma^b \psi
 -\mu \gamma^0 \psi + \frac{1}{4}Q^0 \psi \nonumber \\
 & & -\frac{1}{4}\Omega_{c,ab} \epsilon^{abcd}
     \gamma^0(\gamma_5 \gamma_d) \psi .                \label{h6}
 \ea
Under the definitions $ i\gamma^0 :=-\beta $ and $ \gamma^0
\gamma^i :=-\alpha^i $,
with the constants $\hbar$ and $c$ put in, equation(\ref{h6})
takes the form
 \ba
   i\hbar \frac{\partial \psi}{\partial t} &=& - i\hbar\frac{ fc}{g}
 \alpha^i [\partial_i + \frac{g}{4}Q^i
  + \frac{\partial_i (fg)}{2fg} ] {} \psi  \nonumber \\
 & & + \frac{i\hbar fc}{2} {\Omega^{a,}}_{ab} \gamma^0 \gamma^b \psi
 +fmc^2 \beta \psi
 + \frac{i}{4}\hbar c f Q^0 \psi \nonumber \\
 & & -\frac{i\hbar fc}{4}\Omega_{c,ab} \epsilon^{abcd}
 \gamma^0 (\gamma_5 \gamma_d) \psi  .                    \label{h7}
 \ea
In analogy with the Schr\"{o}dinger equation
 \ba
   i\hbar \frac{\partial \psi}{\partial t} = H \psi
 \ea
we write down the Dirac Hamiltonian matrix as
 \ba
     H &=& \frac{f}{g}c \vec{\pi}.\vec{ \alpha} + fmc^2 \beta
     + \frac{i}{4} \hbar c Q^0
     - \frac{i}{4} \hbar fc Q_i \alpha^i \nonumber \\
     & & + \frac{i\hbar fc}{2} {K^{a,}}_{ab} \gamma^0 \gamma^b
      +\frac{i\hbar fc}{2} {q^{a,}}_{ab} \gamma^0 \gamma^b \nonumber \\
     & & -\frac{i\hbar fc}{4}K_{c,ab} \epsilon^{abcd}\gamma^0 (\gamma_5 \gamma_d)          \label{h8}
 \ea
where
 \ba
 \vec{\pi} :=  \vec{p}
          - \frac{i\hbar}{2fg} \vec{\nabla}(fg) = -i\hbar \vec{\nabla}
          - \frac{i\hbar}{2fg} \vec{\nabla}(fg) .
 \ea
The above expression can be further simplified in terms of the
irreducible components of non-metricity and torsion.
We write
\ba
\imath^{a}Q_{ab} = \Lambda_b + \frac{1}{4} \imath_bQ
\ea
so that
 \ba
   {q^{a,}}_{ab} = \frac{3}{4}Q_b - \Lambda_b \; . \label{h12}
 \ea
From (\ref{D.15}) and defining
 \ba
    \alpha &=& F_c e^c  \; ,\\
    \sigma &=& \frac{1}{3!}\sigma_{abc}e^{abc} \; ,  \\
    t_b &=& \frac{1}{2}t_{ac,b}e^{ac}
 \ea
where $e^{ab} = e^a \wedge e^b $ and so on, we calculate
 \ba
    {K^{a,}}_{ab} &=& {t^a}_{b,a} + F_b \; ,   \label{h14} \\
    K_{[c,ab]} &=& -\frac {1}{6} \sigma_{cab}= \epsilon_{abcf}S^f \; . \label{h15}
 \ea
Here $ S =S^a X_a$ is a vector field.

Now when  (\ref{h12}), (\ref{h14})  and (\ref{h15}) are substituted
in (\ref{h8}) and the identity $ \epsilon_{abcf} \epsilon^{abcd} =-3!\delta^d_f $ is used,  we obtain
 \ba
   H &=& (\frac{f}{g}c \vec{\pi} - \frac{i}{2}f\hbar c \vec{N})
  . \vec{\alpha}  -\frac{3}{2}i\hbar fc S^0 \gamma_5 \nonumber \\
       & & +\frac{i}{2}\hbar fc N^0
          +\frac{3}{2}\hbar fc \vec{S}. \vec{\Sigma} \nonumber \\
        & & + fmc^2\beta                           \label{h16}
 \ea
where
 \ba
     N_b &=& {t^a}_{b,a} +F_b + \frac{5}{4}Q_b
             -\Lambda_b  \; \; ,\\
     \Sigma^i &=&   \left (
                 \begin{array}{cc}
                                   \sigma^i & 0    \\
                                   0    & \sigma^i
                 \end{array}
          \right ).
 \ea
As far as we know, only the cases
involving axial components of torsion had been studied in the literature.
The $ N_a $ terms in the Dirac Hamiltonian
(\ref{h16}) that involve contributions from both the torsion
and non-metricity  are new.

\subsection{ The low energy limit }

If one encounters the situation in which no high momentum states
occur, and the momentum of the Dirac particle remains small
compared with $ mc $ under the influence of all interactions, we
have essentially a non-relativistic problem. It therefore becomes
valid to look at the problem in the context of 2-component wave
functions. The technique for obtaining the Hamiltonian in the
low-energy limit is known as the Foldy-Wouthuysen
transformation~\cite{fol}. Far away from the central gravitating
body of a mass $M$, it is sufficient to consider a weak limit of
the Schwarzschild field (\ref{h1}) with
 \ba
      f \approx 1 + \frac{\vec{g}.\vec{r}}{c^2} \; , \;\; \;\;
      g \approx 1 - \frac{\vec{g}.\vec{r}}{c^2}
 \ea
where
 \ba
      \vec{g}=-GM\frac{\vec{r}}{r^3} \; .
 \ea
Then the Hamiltonian (\ref{h16}) may be written down as
 \ba
    H = mc^2 \beta +\vartheta + \varepsilon        \label{fw1}
 \ea
where
 \ba
    \vartheta = ( \frac{f}{g}c\vec{\pi} -\frac{i}{2}\hbar fc\vec{N})
 .\vec{\alpha} -\frac{3i}{2}\hbar fcS^0 \gamma_5          \label{fw2}
 \ea
is the {\it odd} operator and
 \ba
    \varepsilon = \frac{i}{2}\hbar fcN^0 +
    \frac{3}{2}\hbar fc \vec{S}.\vec{\Sigma}
    + \vec{g}. \vec{r} m \beta  \label{fw3}
 \ea
is the {\it even} operator.

For our case here, distinct from \cite{fol}, in which the odd
terms are removed from the Hamiltonian order by order in $1/m$ and
the even operator does not have the mass term, we have the even
part $\varepsilon $ necessarily containing a term proportional to
the mass $m$. As a result, the removed terms are the same order in
$1/m$ as the kept terms. This makes the issue of convergence of
the approximation scheme problematic. It may be recalled that
there is another approach to the Foldy-Wouthuysen transformation
(see \cite{nik,obu} and references therein). However, it seems
that this approach is not applicable to our case because of the
difficulty of constructing an anticommuting involution operator
(equations(6),(9) in~\cite{nik} and equation(17) in~\cite{obu}).
Thus in spite of the convergence ambiguity we apply the usual
approximation scheme here. Performing the Foldy-Wouthuysen
transformation, we obtain the following Hamiltonian to order
$1/m$,
 \ba
     H = \beta ( mc^2 + \frac{ \vartheta^2 }{2mc^2} ) +
     \varepsilon + \cdots
 \ea
which gives
 \ba
     H &=&  \beta \{ mc^2 + m\beta \vec{g}.\vec{r}
             + \frac{f^2}{2mg^2}(\vec{\pi} )^2 \nonumber \\
     & & { } -\frac{i\hbar f}{2mg}\vec{\nabla}(\frac{f}{g}).\vec{\pi}
             + \frac{\hbar f}{2mg}[\vec{\nabla}(\frac{f}{g})\times
             \vec{\pi}].\vec{\Sigma}    \nonumber \\
      & & { } + \frac{ifc \hbar}{2} N^0
          - \frac{f^2 \hbar^2}{4mg}\vec{\nabla}.\vec{N}
        - \frac{if^2 \hbar^2}{4mg}(\vec{\nabla}\times \vec{N}).
        \vec{\Sigma}  -\frac{i\hbar^2 f}{4mg}(\vec{\nabla} f
       \times \vec{N}).\vec{\Sigma} \nonumber \\
      & & { } -\frac{if^2 \hbar}{2mg}\vec{N}.\vec{\pi}
        + \frac{3if^2 \hbar^2}{4m}S^0\vec{N}.\vec{\Sigma}
      -\frac{\hbar^2 f}{4mg} \vec{N}.\vec{\nabla}f
      - \frac{f^2 \hbar^2}{8m}(\vec{N})^2 \nonumber \\
      & & { } + \frac{3fc\hbar}{2} \vec{S}.\vec{\Sigma}
      - \frac{3f^2 \hbar}{2mg} S^0\vec{\pi}.\vec{\Sigma} \nonumber \\
      & & { } + \frac{3i\hbar^2 f}{4mg} \vec{\nabla}(fS^0).\vec{\Sigma}
         + \frac{9f^2 \hbar^2}{8m}(S^0)^2 \} + \cdots \; . \label{fw4}
 \ea
This Hamiltonian is free of odd operators to order $ 1/m $ and
consequently
to this order its eigenfunctions are 2-component wave functions
corresponding to positive and negative energies. For positive
energies we have the following Pauli Hamiltonian
 \ba
    H=H_0 +H_1 +H_2
 \ea
where
 \ba
     H_0 &=& mc^2 + m\beta \vec{g}.\vec{r}
            + \frac{f^2}{2mg^2}(\vec{\pi} )^2 \nonumber \\
     & &- \frac{i\hbar f}{2mg}\vec{\nabla}(\frac{f}{g}).\vec{\pi}
             + \frac{\hbar f}{2mg}[\vec{\nabla}(\frac{f}{g})\times
             \vec{\pi}].\vec{\sigma} + \cdots    \label{fw5}
 \ea
is the (pseudo-)Riemannian part of the Hamiltonian in the
low-energy limit,
 \ba
     H_1 &=& \frac{3fc\hbar}{2} \vec{S}.\vec{\sigma}
  -\frac{3 f^2 \hbar}{2mg} S^0\vec{\pi}.\vec{\sigma}\nonumber \\
  & &+ \frac{3i\hbar^2 f}{4mg} \vec{\nabla}(fS^0).\vec{\sigma}
         + \frac{9f^2 \hbar^2}{8m}(S^0)^2  + \cdots \label{fw6}
 \ea
contains the axial part of torsion in $H$, and
 \ba
     H_2 &=& \frac{ifc\hbar}{2}N^0
            - \frac{if^2 \hbar}{2mg}\vec{N}.\vec{\pi}
           + \frac{3if^2 \hbar^2}{4m}S^0\vec{N}.\vec{\sigma}
     - \frac{f^2 \hbar^2}{4mg}\vec{\nabla}.\vec{N}
      -\frac{\hbar^2 f}{4mg} \vec{N}.\vec{\nabla}f \nonumber \\
   & & - \frac{if^2\hbar^2}{4mg}(\vec{\nabla}\times \vec{N}).
        \vec{\sigma}  -\frac{i\hbar^2 f}{4mg}(\vec{\nabla} f
       \times \vec{N}).\vec{\sigma}
 - \frac{f^2 \hbar^2}{8m}(\vec{N})^2 + \cdots \label{fw7}
 \ea
contains the other components of torsion and certain components of
non-metricity and constitutes the original content of this work.
Similar calculations have been done in \cite{lam} for effects on
energy levels of atoms coming from  the coupling
just the axial part of torsion to the Dirac equation and in~\cite{ryd}
for the inertial (acceleration and rotation)  and torsional
effects in the low energy limit in the context of
Einstein-Cartan-Dirac theory in which the torsion tensor is purely
axial and in~\cite{obu,ryd2} for the test of equivalence principle
without torsion.

\section{Discussion}

We have studied the motion of a massive, spinning particle in a background
spacetime geometry with torsion and non-metricity.
To this end we have written the Dirac equation in such non-Riemannian spacetimes
including only certain components of non-metricity.
On the other hand all irreducible components of torsion were included for the
 purpose of comparing our results with those already existing in the literature. 
After a Foldy-Wouthuysen transformation the Dirac Hamiltonian
is the sum of three terms where $H_0$ coincides with the original Foldy-Wouthuysen 
expression in the limit $f = g= 1$ and is independent of both torsion and non-metricity. 
$H_1$ depends only on the axial components
of the torsion tensor while the last term $H_2$
is proportional to $N_a$'s that involve both torsion and non-metricity
components. All the contributions to $H_2$ turn out to be of the order of $\hbar$ or of higher orders.
Consequently their effect would be too small to be macroscopically significant. 
However, at the microscopic scales the coupling of a Dirac particle to spacetime 
non-metricity might have some interesting implications.

\bigskip

\noindent {\Large {\bf Acknowledgement}}

\medskip

\noindent This work has been supported in part by T\"{U}B\.{I}TAK
(The Scientific and Technical Research Council of Turkey).
We thank Professor R. W. Tucker and Dr. P. Nurowski for discussions.

\end{document}